\documentclass[12pt]{article}

\usepackage{amsmath,amssymb,amsfonts,amsthm,dsfont,dcolumn}
\usepackage[margin=-5pt]{caption}
\usepackage{graphicx,wrapfig}
\usepackage{tensor}
\usepackage{appendix}

\DeclareGraphicsExtensions{.pdf,.png,.jpg,.jpeg}

\newcolumntype{d}{D{.}{.}{0}}

\newlength{\textlength}
\newlength{\overlinelength}

\usepackage{color}

\newcommand{\ket}[1]{\left| #1 \right>} 
\newcommand{\bra}[1]{\left< #1 \right|} 
\newcommand{\braket}[2]{\left< #1 \vphantom{#2} \right|
 \left. #2 \vphantom{#1} \right>} 

\newcounter{subequation}[equation]

\newcommand{\be}{\begin{equation}}
\newcommand{\ee}{\end{equation}}
\newcommand{\eel}[1]{\label{#1}\end{equation}}
\newcommand{\bea}{\begin{eqnarray}}
\newcommand{\eea}{\end{eqnarray}}
\newcommand{\eeal}[1]{\label{#1}\end{eqnarray}}

\makeatletter

\def\thesubequation{\theequation\@alph\c@subequation}
\def\@subeqnnum{{\rm (\thesubequation)}}
\def\slabel#1{\@bsphack\if@filesw {\let\thepage\relax
   \xdef\@gtempa{\write\@auxout{\string
      \newlabel{#1}{{\thesubequation}{\thepage}}}}}\@gtempa
   \if@nobreak \ifvmode\nobreak\fi\fi\fi\@esphack}
\def\subeqnarray{\stepcounter{equation}
\let\@currentlabel=\theequation\global\c@subequation\@ne
\global\@eqnswtrue \global\@eqcnt\z@\tabskip\@centering\let\\=\@subeqncr

$$\halign to \displaywidth\bgroup\@eqnsel\hskip\@centering
  $\displaystyle\tabskip\z@{##}$&\global\@eqcnt\@ne
  \hskip 2\arraycolsep \hfil${##}$\hfil
  &\global\@eqcnt\tw@ \hskip 2\arraycolsep
  $\displaystyle\tabskip\z@{##}$\hfil
   \tabskip\@centering&\llap{##}\tabskip\z@\cr}
\def\endsubeqnarray{\@@subeqncr\egroup
                     $$\global\@ignoretrue}
\def\@subeqncr{{\ifnum0=`}\fi\@ifstar{\global\@eqpen\@M
    \@ysubeqncr}{\global\@eqpen\interdisplaylinepenalty \@ysubeqncr}}
\def\@ysubeqncr{\@ifnextchar [{\@xsubeqncr}{\@xsubeqncr[\z@]}}
\def\@xsubeqncr[#1]{\ifnum0=`{\fi}\@@subeqncr
   \noalign{\penalty\@eqpen\vskip\jot\vskip #1\relax}}
\def\@@subeqncr{\let\@tempa\relax
    \ifcase\@eqcnt \def\@tempa{& & &}\or \def\@tempa{& &}
      \else \def\@tempa{&}\fi
     \@tempa \if@eqnsw\@subeqnnum\refstepcounter{subequation}\fi
     \global\@eqnswtrue\global\@eqcnt\z@\cr}
\let\@ssubeqncr=\@subeqncr
\@namedef{subeqnarray*}{\def\@subeqncr{\nonumber\@ssubeqncr}\subeqnarray}

\@namedef{endsubeqnarray*}{\global\advance\c@equation\m@ne
                           \nonumber\endsubeqnarray}

\makeatletter \@addtoreset{equation}{section} \makeatother
\renewcommand{\theequation}{\thesection.\arabic{equation}}

\catcode`\@=11

\newcount\hour
\newcount\minute
\newtoks\amorpm \hour=\time\divide\hour by 60\minute
=\time{\multiply\hour by 60 \global\advance\minute by-\hour}
\edef\standardtime{{\ifnum\hour<12 \global\amorpm={am}
        \else\global\amorpm={pm}\advance\hour by-12 \fi
        \ifnum\hour=0 \hour=12 \fi
        \number\hour:\ifnum\minute<10
        0\fi\number\minute\the\amorpm}}
\edef\militarytime{\number\hour:\ifnum\minute<10 0\fi\number\minute}

\def\draftlabel#1{{\@bsphack\if@filesw {\let\thepage\relax
   \xdef\@gtempa{\write\@auxout{\string
      \newlabel{#1}{{\@currentlabel}{\thepage}}}}}\@gtempa
   \if@nobreak \ifvmode\nobreak\fi\fi\fi\@esphack}
        \gdef\@eqnlabel{#1}}
\def\@eqnlabel{}
\def\@vacuum{}
\def\marginnote#1{}
\def\draftmarginnote#1{\marginpar{\raggedright\scriptsize\tt#1}}
\def\draft{
        \pagestyle{plain}
        \overfullrule=2pt
        \oddsidemargin -.5truein
        \def\@oddhead{\sl \phantom{\today\quad\militarytime} \hfil
        \smash{\Large\sl DRAFT} \hfil \today\quad\militarytime}
        \let\@evenhead\@oddhead
        \let\label=\draftlabel
        \let\marginnote=\draftmarginnote
        \def\ps@empty{\let\@mkboth\@gobbletwo
        \def\@oddfoot{\hfil \smash{\Large\sl DRAFT} \hfil}
        \let\@evenfoot\@oddhead}

\def\@eqnnum{(\theequation)\rlap{\kern\marginparsep\tt\@eqnlabel}
        \global\let\@eqnlabel\@vacuum}  }

\renewcommand{\theequation}{\thesection.\arabic{equation}}

\def\appendix#1{
  \addtocounter{section}{-3}
  \setcounter{equation}{0}
  \renewcommand{\thesection}{\Alph{section}}
  \section*{Appendix \thesection\protect\indent \parbox[t]{11.15cm}
  {#1} }
  \addcontentsline{toc}{section}{Appendix \thesection\ \ \ #1}
  }
\def\be{\begin{equation}}
\def\ee{\end{equation}}

\date{}
\begin{document}

\begin{titlepage}
\hfill MCTP-14-17\\

\begin{center}

{\Large \bf Left-Right Entanglement Entropy of Boundary States}

\vskip .7 cm

\vskip 1 cm

{\large   Leopoldo A. Pando Zayas$^1$ and Norma Quiroz$^2$}

\end{center}

\vskip .4cm \centerline{\it ${}^1$ Michigan Center for Theoretical
Physics}
\centerline{ \it Randall Laboratory of Physics, The University of
Michigan}
\centerline{\it Ann Arbor, MI 48109-1120}

\vskip .4cm \centerline{\it ${}^2$  Facultad de Ciencias}
\centerline{ \it Universidad de Colima}
\centerline{\it Bernal D\'{\i}az del Castillo 340, Col. Villas San Sebasti\'an,
Colima}
\centerline{\it Colima 28045, M\'exico}

\vskip .4cm
\centerline{ \it }
\centerline{\it  }

\vskip 1 cm

\vskip 1.5 cm

\begin{abstract}
We study entanglement entropy of boundary states in a free bosonic conformal field theory. A boundary state can be thought of as composed of a particular combination of left and right-moving modes of the two-dimensional conformal field theory. We investigate the reduced density matrix obtained by tracing over the right-moving modes in various boundary states.  We consider Dirichlet and  Neumann boundary states of a free noncompact as well as a compact boson. The results for the entanglement entropy indicate that the reduced system can be viewed as a thermal CFT gas. Our findings are in agreement and generalize results in quantum mechanics and quantum field theory where coherent states can also be considered. In the compact case we verify that the entanglement entropy expressions are consistent with  T-duality.
\end{abstract}

\end{titlepage}


\date{}



\section{Introduction}

Entanglement entropy has emerged as an important quantitative measure of the entanglement between parts of a system and it plays a particularly important role in identifying subtle phases in some condensed matter system. The traditional paradigm for entanglement entropy assumes a bi-partition of a system into subsystems $A$ and $B$. The entanglement entropy of a region, $A$, is given by the von Neumann entropy formula for the reduced density matrix, $\rho_A$, that is, for the density matrix obtained by tracing over the degrees of freedom of the subsystem $B$, $\rho_A=\mbox{Tr}_B \, \rho$ . Typically, subsystems $A$ and $B$ are defined as geometric regions of the whole system. For example, in two-dimensional conformal field theories (CFT) a typical choice of region $A$ is an interval or a combination of disjoint intervals (see \cite{Calabrese:2004eu} and  \cite{Calabrese:2005zw} for reviews). In this paper we will deviate from a precise geometric prescription for the subsystems and, instead, defined them in terms of left- or right-moving modes.

The original studies of entanglement entropy focused on systems in their ground state \cite{Bombelli:1986rw,Srednicki:1993im}. Once the entanglement entropy of a subsystem which forms part of a system in its ground state is understood, a natural question is to study the properties when the whole system is in an excited state (see for example \cite{Alcaraz:2011tn}).
The interests in entanglement entropy has recently expanded to 	practitioners of string theory thanks to the pioneering work of Ryu and Takayanagi who proposed a connection between entanglement entropy and gravity \cite{Ryu:2006bv,Ryu:2006ef} in the context of the AdS/CFT correspondence. This development, subsequently, allows to understand a first-law like relationship for the entanglement entropy and the energy of excited states geometrically; this connection arises as an implication of Einstein equations \cite{Bhattacharya:2012mi}. Note, however, that such relations can be obtained purely in field theoretic terms \cite{Blanco:2013joa}, \cite{Wong:2013gua} but the gravitational picture anticipated their existence naturally. There have also been some recent work discussing excited states in CFT's in various dimensions \cite{Nozaki:2014hna,Caputa:2014vaa}.

Entanglement entropy has thoroughly been studied in the context of two-dimensional CFT (see reviews \cite{Calabrese:2004eu,Calabrese:2005zw}). In 2D CFT the entanglement entropy  grows with the logarithm of the subsystem size.
If the 2D CFT is defined on a manifold with a boundary, then there exist states called boundary states which are  similar to the coherent state of coupled harmonic oscillators.  The boundary state admits a Schmidt decomposition with respect to left and right modes of the Hamiltonian. This decomposition precisely exhibits the boundary state as an entangled stated of left and right modes.

In this  letter we explore the entanglement and R\'enyi entropies of boundary states in boundary conformal field theories (BCFT). Boundary states in BCFT are generalizations of coherent states which are typical in quantum mechanics and quantum field theory.  Various results have been obtained in the context of condensed matter  \cite{PhysRevB.88.075112}, including also applications to the entanglement of topologically ordered states \cite{2012PhRvL.108s6402Q}.

Our main motivation for studying boundary states stems from their central role in string theory.  It is hard to overstate the role of D-branes in string theory; they have provided an invaluable window into nonperturbative aspects of string theory. More importantly, their dynamics underlies entire fields such as AdS/CFT and string phenomenology. In this paper we study the entanglement and R\'enyi entropies of boundary states for the free and compact boson CFT by integrating over the right-moving degrees of freedom.

The paper is organized as follows. In section \ref{Sec:Harmonic} we review the simple case of coupled harmonic oscillators as motivation to our main investigation. In section \ref{Sec:EEBS} we consider the entanglement entropy of various boundary states. This is the main section of the paper and we consider Dirichlet and  Newmann boundary conditions as well a boundary states on a compact space. We conclude in section \ref{Sec:Conclusions} and relegate a number of notational and technical details to an appendix.

\section{Prelude: Entangled harmonic oscillators}\label{Sec:Harmonic}

Entanglement is inherently associated with composite quantum systems and quantifies, at some level, the non-local quantum correlations between the different parts of the  composite system.

There are various results regarding entanglement entropy in the geometric setup, that is, when one of the subsystems is defined by a geometric region. Since we are interested in understanding entanglement entropy in a non-geometric setup, the best way for us to develop our intuition would be to first consider various situations in the case of coupled harmonic oscillators.

In order to setup our notation,  let us consider a pure state $\ket{\psi}$ describing a composite system which is divided into two subsystems $A$ and $B$. The division between the subsystems can be in configuration space or in the Hilbert space. In our case we consider the Hilbert space factorized as ${\cal H} = {\cal H}_A \otimes {\cal H}_B$.
By the Schmidt  decomposition, there exists orthonormal basis $\ket{i_A}$ and $\ket{i_B}$ for subsystems $A$ and $B$ respectively such that $\ket{\psi}=\sum_i \lambda_i \ket{i_A} \ket{i_B}$ where
 $\lambda_i$ are non-negative real numbers (the Schmidt numbers) satisfying  $\sum_i \lambda_i^2 = 1$. In quantum mechanical systems entanglement is encoded in the structure of $\{\lambda_i\}$. For instance, if  one of the $\lambda_i$ is one and the rest are zero, the state $\ket{\psi}$  is a product state, however if all the $\lambda_i$ are the same, the state is maximally entangled.

In order to quantify entanglement we recall that a pure state can also be described by the density operator  $\rho=\ket{\psi}\bra{\psi}$.  This density operator has to satisfy conservation of probability: $\mbox{Tr} \rho =1$.  The reduced density matrix for the subsystem $A$ is defined as $\rho_A=\mbox{Tr}_B \, \rho$, here $\mbox{Tr}_B$ is the partial trace over the subsystem $B$. This definition enables us to compute expectation values as $\langle {\cal O} \rangle = \mbox{Tr} \rho_A {\cal O}$ for all observable measurements within the subsystem $A$.   If the total density matrix, $\rho$, describes a pure state, it  is not in general true that $\rho_A$, obtained by the partial trace, also describes a  pure state. Only when the composite system is in a product state the density matrix $\rho_A$ describes also a pure state.

There are several quantities for measuring entanglement between the subsystems. We will be mostly interested in the entanglement and the R\'enyi entropies. Both of these entropies are defined in terms of the reduced density matrix $\rho_A$.
The entanglement entropy is given by the von Neumann entropy $S= - \mbox{Tr}\, \rho_A \mbox{ln} \rho_A$ while
the R\'enyi entropy is defined as $S_n= \frac{1}{1-n}\mbox{ln}\,\mbox{Tr}\rho^n_A$ for non-negative $n$ different from one. The entanglement entropy can be obtained  from the R\'enyi entropy in the limit $n$ goes to 1.

To develop our intuition about integrating over modes, let us consider a simple composite system formed by two coupled harmonic oscillators \cite{Bombelli:1986rw,Srednicki:1993im}.
The Hamiltonian of such a system is:
\be
\label{Eq:Hamiltonian}
H=\frac{1}{2}\left[ p_1^2 + p_2^2 + k_0(x_1^2 + x_2^2)+k_1(x_1-x_2)^2     \right].
\ee
As can be seen above, the Hamiltonian contains an interaction term. However, using the normal-modes coordinates, the Hamiltonian can be factorized as the sum of two Hamiltonians of two  independent harmonic oscillators. In such coordinates the ground state of the system is the product of two Gaussian states. Namely,
\be
\Psi_{G.S.}(x_1,x_2)=\sqrt[4]{\frac{w_+ w_-}{\pi^2}}\mbox{exp}\left(
-\frac{1}{2}(w_+x_+^2 + w_- x_-^2) \right)
\ee
where $x_\pm=(x_1\pm x_2)/\sqrt{2}$, $w_+=\sqrt{k_0}$ and $w_-=\sqrt{k_0+2k_1}$. This wave function can be expressed as
\be
\label{eq:gaussian}
\ket{\psi}=\frac{1}{\mbox{cosh}\eta}\sum_{n=0}^\infty (-\mbox{tanh}\eta)^n \ket{\phi_n(x_1)}_A\ket{\phi_n(x_2)}_B,
\ee
with $e^{4\eta}= w_-/w_+$ and $\ket{\phi_n}$ the state of a harmonic oscillator of frecuency $w=\sqrt{w_+w_-}$ \cite{2012BrJPh..42..267P}.  The subsystem A corresponds to the first oscillator and  the subsystem B to the second one.
This is a pure state that is already in the Schmidt decomposition in the factorized Hilbert space
${\cal H}={\cal H}_{osc1}\otimes {\cal H}_{osc2}$.
The reduced density matrix is
obtained by tracing out over the second  oscillator

\be
\rho_A= \mbox{Tr}_B\rho=\frac{1}{\mbox{cosh}^2 \eta}\sum_{m=0}^{\infty}(\mbox{tanh}\eta)^{2m}\ket{\phi_m(x_1)}_A\bra{\phi_m(x'_1)}.
\ee

The entanglement entropy is then
\cite{Bombelli:1986rw,Srednicki:1993im}:
\be
\label{eq:tentropy}
S= - \mbox{ln}(1-\gamma^2)-\frac{\gamma^2 \mbox{ln}\gamma^2}{1-\gamma^2}.
\ee
where $\gamma=\mbox{tanh}\eta$. Therefore,  for  $\gamma= \mbox{exp}(-w/T)$ the entanglement entropy (\ref{eq:tentropy}) is interpreted as the entropy of a thermal oscillator\footnote{In units where  $\hslash=k_B=1$.}.  This result is in precise agreement with our intuition that integrating over degrees of freedoms in a system is akin to putting the system in a thermal bath.

\subsection{Coupled harmonic oscillators: Beyond the ground state}

Having seen that integrating over one oscillator leads to thermal behavior when the original system is in the ground state, a very natural question is to revisit this intuition when the original system is in an excited state. First, one would like to understand some simple excited states such as: low excited states and eventually coherent states. For the case of harmonic oscillators some low excited states and their entanglement entropy were considered in \cite{Ahmadi:2005mh,Das:2005ah}; our interest is centered on coherent states.


In the case of quantum mechanics, a coherent state of a harmonic oscillator can be defined as  an eigenstate of the annihilation operator. Namely, $\ket{\alpha}$ such that,

\be
\hat{a}\ket{\alpha}=\alpha \ket{\alpha}.
\ee
A solution is generally of the form
\be
\ket{\alpha}=\exp(\alpha \hat{a}^{\dagger}-\alpha^* a)\ket{0}=\exp\left(-\frac{|\alpha|^2}{2}\alpha \hat{a}^{\dagger}\right)\ket{0}.
\ee

Since these states are eigenstates of the momentum operator we can view them as simply shifting the ground state. In the context of the Hamiltonian Eq. \ref{Eq:Hamiltonian}, one can consider the entanglement entropy on a coherent state after tracing over one of the harmonic oscillators. The key observation in the computations comes from the simplification given by the fact that the coherent state wave function of two-coupled harmonic oscillators, written in terms of normal-mode coordinates ($x_{\pm}$), can be expressed as a product of two Gaussians representing the ground states of  individual oscillators
\be
{\psi_{CS}(x_1,x_2)}=\psi_0(x_+ -a)\psi_0(x_- -b).
\ee
where the constants  $a$ and $b$ are complex numbers denoting the displacements of the ground states in the phase space. This wave function can be expanded as in equation (\ref{eq:gaussian}).  Then  the  entanglement entropy  is given by Eq.(\ref{eq:tentropy}). Therefore, the  entanglement entropy for general coherent states of two coupled harmonic oscillators is the same as that of the ground state \cite{Ahmadi:2005mh,Das:2005ah}, quite surprisingly.

It is possible to argue, {\it a posteriori}, that the coherent states are, in the sense of the uncertainty principle, very classical and, therefore, do not contribute to quantum entanglement beyond that already present in the vacuum state. We will see that a modified version of this statement holds true for boundary states in 2d CFT.

\section{Entanglement entropy for boundary states of  free boson}\label{Sec:EEBS}

Conformal field theory plays an important role in string theory where it arises as a two-dimensional field theory on the world-sheet of the string. A key role within string theory is played by D-branes whose conformal field theory realization is in terms of boundary states.

From the spacetime point of view, D-branes are extended objects where the endpoints of the open strings can end.  D-branes play a fundamental role in the formulation of non-perturbative string theory and are essential in string phenomenology.
When the open string moves on the 10D space-time, it spans an infinite strip, with spatial coordinate $\sigma \in [0,\pi]$ and temporal coordinate $\tau \in (-\infty, \infty)$. An alternative view is provided by first defining the complex coordinate $w= \sigma + i \tau$ and then performing the conformal map $z=e^{-iw}$. Under this map,  the infinite strip is transformed to the upper-half plane  ($\mbox{Im}z\geq 0$) and the end points of the open strings are mapped to the real axis ($\mbox{Im}z = 0$). Therefore, from the point of view of the two-dimensional world-sheet, the D-brane is described by a boundary  conformal field theory (BCFT), that is, a CFT with extra data given at its boundary ($\mbox{Im}z = 0$).
Different D-branes  are defined by the boundary conditions the bosonic space-time coordinates $X(z,\bar{z})$ satisfy on the real axis.

There are two ways of describing the interactions between D-branes: the open and closed string channels. In the open string channel an interaction between two D-branes is given by two open strings stretched between two D-branes and interacting at one-loop.
The one-loop diagram  is  a cylinder  with periodic time
$\tau \sim  \tau + 2 \pi t$ where $t \in (-\infty, \infty)$ is the modular parameter of the cylinder. Such process
 is described by a one-loop  partition function $Z_o (t)=\mbox{Tr  exp}(-2\pi t H_o)$ where $H_o$ is the open string Hamiltonian.

One can switch from the open string picture to the closed string channel by performing a conformal transformation which interchanges the roles of $\tau$ and $\sigma$. Under this modular transformation, the interaction in the closed string picture is described by a cylinder diagram with periodic spatial coordinate $\sigma \sim \sigma + 2\pi$ and time along the cylinder of length $2 \pi l$. The tree-level amplitude is $Z_c (l) = \bra{B}\mbox{exp}(-2\pi l H_c)\ket{B}$, where $H_c$ is the closed string Hamiltonian (See Appendix for these definitions). This
describes the process of a closed string that is emitted at a boundary state $\ket{B}_1$ then propagates to other boundary state $\ket{B}_2$ and it is absorbed there.  These two forms for describing the interaction between branes are equivalent and the amplitudes in the open and closed string channels are the same. This requirement is some times called the {\it loop-channel-tree-channel equivalence} and it fixes $t=1/2l$.\\
In this context, the open string boundary conditions can be described in terms of boundary conditions for boundary states $\ket{B}$. For the boundary located at $\tau=0$ the  possibilities are:
\be
\label{eq:Nbc}
 \partial_\tau X(\tau,\sigma)\ket{B}=0,   \hspace{2cm} \mbox{Neumann boundary condition}
 \ee

 and

 \be
 \label{eq:Dbc}
 \partial_\sigma X(\tau,\sigma)\ket{B}=0,   \hspace{2cm} \mbox{Dirichlet boundary condition}
 \ee
 where $X(\tau,\sigma)$ is the closed string field that maps the worldsheet to the spacetime and is defined in the appendix.

 \subsection{Entanglement entropy for Neumann boundary state}
Expanding $X(\tau,\sigma)$ in terms of Laurent modes the boundary condition (\ref{eq:Nbc}) is expressed as
\be
(\alpha_n + \tilde \alpha_{-n}) \ket{B}=0, \qquad  p\ket{B}=0,
\ee
for any integer $n$ and $\alpha_n$ and $\tilde \alpha_n$ are the left (L) and right (R) oscillator modes respectively.  The relation $p=0$ express the usual condition of vanishing momentum flow through the boundary.
 A solution to this equation is
\be
\label{eq:Nbs}
\ket{B}={\cal N}{\mbox {exp}}\Bigl(- \sum_{k=1}^\infty \frac{1}{k}\alpha_{-k}{\tilde \alpha_{-k}} \Bigr)\ket{0},
\ee
the normalization is
determined by the {\it loop-channel-tree-channel equivalence}. Equation (\ref{eq:Nbs}) can be expressed as \cite{Ishibashi:1988kg,Blumenhagen:2009zz}
\be
\label{eq:Nbsent}
\ket{B}={\cal N}\sum_{\vec m}\ket{\vec m} \otimes \ket{U \tilde{\vec{m}}},
\ee
where  the states

\begin{align}
\ket{\vec{m}}&= \ket{m_1,m_2,\ldots}=\prod_{t=1}^{\infty}\frac{1}{\sqrt{m_t!}}
\left(\frac{\alpha_{-t}}{\sqrt{t}}\right)^{m_t}\ket{0}, \nonumber \\
\ket{\vec{\tilde m}}&= \ket{\tilde m_1,\tilde m_2,\ldots}=\prod_{t=1}^{\infty}\frac{1}{\sqrt{m_t!}}
  \left(\frac{\tilde \alpha_{-t}}{\sqrt{t}}\right)^{m_t}\ket{0}\,,
\end{align}
are a complete orthonormal basis for ${\cal H}_L$ and ${\cal H}_R$, respectively and $U$ is an anti-unitary operator acting on Hilbert space  ${\cal H}_R$.

The form of Eq. \ref{eq:Nbs} allows us to identify the boundary state as a coherent state in the Hilbert space ${\cal H}_L \otimes {\cal H}_R$ while Eq. \ref{eq:Nbsent} could be thought of as
 a Schmidt decomposition of the boundary state  with all the coefficients of the decomposition the same ($\lambda_i = {\cal N}$), thus expressing maximal entanglement between left and right modes.  Similar maximally entangled states have been discussed recently in \cite{Nozaki:2014hna,Caputa:2014vaa}.

Given a coherent state $\ket{B}$ it is natural, in the context we are discussing, to study the associate density matrix, $\rho=\ket{B}\bra{B}$.
However, the above na\"ive definition of $\rho$ does not satisfy  $\mbox{Tr}\rho =1$. The main culprit is that the standard inner product $\braket{B}{B}$ diverges \cite{Ishibashi:1988kg} as  can be seen from the limit $l \rightarrow 0$ of the tree-level amplitude $Z_c(l)$.

More precisely, the conformal boundary state $\ket{B}$ is scale invariant and that is the main reason for its non-normalizability  \cite{Ishibashi:1988kg}. To remedy this situation we will follow a fairly established prescription first introduced, to the best of our knowledge, by Takayanagi and Ugajin in \cite{Takayanagi:2010wp}. This prescription has recently been used  by Cardy in \cite{Cardy:2014rqa} and consists in  introducing a finite correlation length by considering instead: $e^{-\epsilon H}\ket{B}$.  This choice has been further argued for in \cite{Calabrese:2006rx,Calabrese:2007rg} on phenomenological grounds connected to quantum quenches and by representing an action on $\ket{B}$ with the most irrelevant operator\footnote{In the applications described in \cite{Calabrese:2006rx,Calabrese:2007rg}, the state has been denoted by $e^{-(\beta/4) H}\ket{B}$ as corresponding to a quench.}. We define, therefore,  the density matrix, accordingly as
\be\label{eq:regularized}
\rho = \frac{e^{-\epsilon H}\ket{B}\bra{B}e^{- \epsilon H}} {\cal{A}},
\ee
where ${\cal A}$ is a constant that is fixed by requiring $\mbox{Tr} \rho =1$. The Hamiltonian,  $H$, is the closed string Hamiltonian described in the appendix.

A expression similar to Eq. \ref{eq:regularized}, has been studied in  \cite{Takayanagi:2010wp,Cardy:2014rqa} in the context of quantum quenches  and in \cite{Nozaki:2014hna} for excited states defined by local operators. Another way of thinking of $\epsilon$, as originally suggested in \cite{Takayanagi:2010wp} is to view it as parametrizing a UV filter. In this case if the UV cutoff or lattice spacing of the field theory is $a_{UV}$, one would need $\epsilon \ll a_{UV}$.  Alternatively, one can think of $\epsilon$ as used to infinitesimally   evolved the state $|B\rangle$ using the Hamiltonian $H$. This evolution allows to make $\braket{B}{B}$ formally finite thus allowing us to correctly normalize the density matrix corresponding to the evolved $|B\rangle$.

The reduced density matrix $\rho_A = {\mbox Tr}_{B} \rho$, which is made on the right-modes (system B), is given in Eq.(\ref{eq:rA}). The corresponding   von Neumann entropy is
\be
S_A= \mbox{ln}Z - \mbox{ln}q \frac{d}{d\mbox{ln}q}\mbox{ln}Z,
\ee
where $Z=\prod_{l=1} \frac{1}{1-q^l}$ with $q= e^{-8\pi \epsilon}$. We recognize $Z$ as the partition function of a gas of photons if $\beta  \omega_0= 8\pi \epsilon$\footnote{Recall that the partition function of a gas of photons is given by $Z=\sum \exp(-\beta n\hbar \omega_o)=1/(1-e^{-\beta\hbar\omega_0})$. Moreover, the full partition function is then obtained by multiplying the result for one frequency mode over all possible frequencies.}.  In such case the entropy reduces to the thermodynamical entropy
\be
\label{eq:vNE}
S_A=\frac{\pi}{24\epsilon}+\ldots \,.
\ee
This entropy was already recognized in \cite{Takayanagi:2010wp} as the entropy corresponding to a thermal CFT gas at the effective temperature $T_{eff}=1/4\epsilon$. Note that this is precisely compatible with the interpretation of Cardy in \cite{Cardy:2014rqa}.
To get more insight into the system A, we compute trace of $\rho_A^n$ which is given as

\be\label{eq:tracen}
\mbox{Tr}\rho_A^n \propto \frac{Z_c(2\epsilon n)}{Z_c(2 \epsilon)^n},
\ee
where $Z_c(2 \epsilon n)$ is the tree-level amplitude (in the closed channel) on circle propagating between the two boundaries along an imaginary time $2 \epsilon n$. The proportionality constant is irrelevant since it cancels when expressing  $Z_c$ in terms of the Dedekind  eta function.
For $\epsilon \rightarrow 0$, the factor $q=e^{-8\pi \epsilon}$ approaches to unit. To further proceed, we use the {\it loop-channel-tree-channel equivalence} to express $Z_c(2 \epsilon n)$  as a partition function in the open string channel  with the factor $\tilde{q}=\exp(-\frac{\pi}{2\epsilon n})$.
As $\epsilon$ goes to zero $\tilde q$ goes to zero, so the leading term is the ground state. In such case
\be
\mbox{Tr}\rho_A^{(n)}\sim 2^{1-n}\sqrt{n}\epsilon^{\frac{1}{2}(1-n)} e^{\frac{1}{24}\frac{\pi}{2\epsilon}(\frac{1}{n}-n)}\,.
\ee
The entanglement entropy is given by the limit when $n$ goes to one of the R\'enyi entropy and one gets:
\be
\label{eq:VNentropy}
S_A= \frac{\pi}{24\epsilon}+\mbox{ln}2-\frac{1}{2} + \frac{1}{2}\mbox{ln}\epsilon
\ee
This way we have managed to obtain some corrections to the leading term in expression Eq. \ref{eq:vNE}. The second term in the above expression Eq. \ref{eq:VNentropy} is the boundary entropy of the boundary state defined in Sec.\ref{Sec:app} with negative sign \cite{1751-8121-42-50-504009,Cardy:2004hm}. Although the boundary entropy is generically derived in the limit of $\epsilon \to \infty$ (see \cite{Affleck:1991tk} and a detailed discussion in \cite{Blumenhagen:2009zz}), it is plausible that modular invariance brings back such contribution.

Let us comment on the leading term. Normally, the divergences in the entanglement entropy are attributed to arbitrarily high energy correlations between nearby degrees of freedom on the two sides of the curve dividing  the subsystems $A$ and $B$. Since we are not geometrically separating subsystems $A$ and $B$ we can not claim those correlations as the source of our divergence. However, it could be argued, as in most field theories, that the divergence com from allowing arbitrarily high energy degrees of freedom to be traced over. Namely, we can track the divergence to the fact that we sum over all modes $\tilde{\alpha}_n$ which have increasingly higher and higher energy.

\subsection{Entanglement entropy for Dirichlet boundary state}

The other possibility for boundary conditions on the end points of the open string is the Dirichlet condition. When expressed in terms of  the oscillator modes of the closed string as:
\be
(\alpha_n - \tilde \alpha_{-n}) \ket{B}=0, \qquad  x\ket{B}=a\ket{B},
\ee
where $x$ is the center of mass position of the closed string. According to the  equation on the right the Dirichlet boundary  state carries arbitrary momentum $p$.
The solution is
\be
\label{eq:Dbs}
\ket{B,p}={\cal N} \exp\Bigl( \sum_{k=1}^\infty \frac{1}{k}\alpha_{-k}{\tilde \alpha_{-k}} \Bigr)\ket{0}\ket{p}.
\ee
After a Fourier transformation, the boundary state localized in $a$ is:
\be
\ket{B, a}=\int dp\, e^{i p a}\ket{B,p}.
\ee

The density matrix has now the form
\be
\rho = \frac{1}{{\cal A}} e^{-\epsilon H}\ket{B,a}\bra{B,a}e^{-\epsilon H}.
\ee
The trace of $\rho_A^n$ is given in Eq.(\ref{eq:Dtracen}), and because of the {\it open-closed equivalence} one gets in the limit $\epsilon$ small
\be
\mbox{Tr}\rho^n_A \sim e^{\frac{1}{24}\frac{\pi }{2\epsilon}(\frac{1}{n}-n)}\,.
\ee
The contribution of the zero modes to the boundary state cancel the constant terms and factors powers of  $\epsilon$ present in the Neumann case; a similar behavior was observed in \cite{PhysRevB.88.075112}.  Hence the entanglement entropy is only thermal. There is no contribution of the boundary entropy as expected since the normalization constant of the Dirichlet brane is one.

\subsection{Entanglement Entropy  for boundary state of a compact boson}
We now consider the boundary state when the spatial dimension is compactified on a circle of radio $R$. As an effect of such compactification, the left and right zero modes of the closed string field  are not in general the same and their eigenvalues take discrete values.

For Neumann boundary conditions the boundary state now  satisfies:
$$
\left( p_L+p_R \right)\ket{B}=0,\;\;\;\;\;\; \left(\alpha_n + \tilde \alpha_{-n}\right)\ket{B}=0.
$$
The first condition fix $p_L=-p_R$ therefore  $m=0$ and the Neumann boundary state has only winding modes $k$ and a  Wilson line $w$.
\be
\ket{B(w)}= {\cal N}\sum_{k}e^{i\frac{wkR}{2}} e^{- \frac{1}{l}\sum_{l}\alpha_{-l}{\tilde \alpha}_{-l}}
\ket{\frac{kR}{2},\frac{-kR}{2}}.
\ee
The density matrix is defined as equation (\ref{eq:regularized}). The result for the R\'enyi is computed in appendix. After the open-closed equivalence and in the limit for $\epsilon$ going to zero
$$
\mbox{Tr}\rho_A^n \sim 2^{1-n}R^{n-1} e^{\frac{1}{24}\frac{\pi}{2\epsilon}(\frac{1}{n}-n)}.
$$
From here
\be
\label{eq:cnentropy}
S^N_A = \mbox{ln}2 - \mbox{ln}R + \frac{\pi}{24 \epsilon}.
\ee
For Dirichlet boundary conditions the boundary state satisfy
$$
\left( p_L-p_R \right)\ket{B}=0,\;\;\;\;\;\; \left(\alpha_n - \tilde \alpha_{-n}\right)\ket{B}=0\,.
$$
The first condition restrict $p_L=p_R$ therefore $k=0$ and the boundary state carries only momentum along the compact direction. A solution is the boundary state at  position $a$
\be
\ket{B,a}= \sum_m e^{i\frac{ma}{R}} e^{ \sum_{k=1}^\infty \frac{1}{k}\alpha_{-k}{\tilde \alpha_{-k}}} \ket{\frac{m}{R},\frac{m}{R}}.
\ee

The density matrix is defined as before and by the same procedure as above one gets

$$
\mbox{Tr}\rho_A^n \sim R^{1-n} e^{\frac{1}{24}\frac{\pi}{2\epsilon}(\frac{1}{n}-n)}\,,
$$
then

\be\label{eq:cdentropy}
S^D_A =  \mbox{ln}R + \frac{\pi}{24 \epsilon}\,.
\ee
We can see that equations (\ref{eq:cnentropy}) and (\ref{eq:cdentropy}) are compatible with T-duality: $R \rightarrow \frac{2}{R}$. The results have some resemblance to the calculations presented in \cite{Hsu:2008af}, even though the context is quite different.
\section{Conclusions}\label{Sec:Conclusions}

In this paper we have investigated the effects of tracing over the left-moving degrees of freedom of various boundary states. In particular we have considered Dirichlet and  Neumann boundary states of a free boson CFT. We have also performed similar calculations for the compact boson. In the latter case we verify that the entanglement entropy is compatible with  T-duality.

Our choice of division of the system into subsystems $A$ and $B$ does not follow the more traditional geometric delimitation. Rather, we have decomposed the Hilbert space into left- and right-moving degrees of freedom. Our approach generalized the intuition developed in quantum mechanics of tracing over a set of harmonic oscillators for a system of many coupled oscillators. This approach is, perhaps, more akin to recent investigation of momentum entanglement \cite{Balasubramanian:2011wt}, \cite{Hsu:2012gk}, \cite{Lundgren:2014qua}. However, it is quite natural, in the context of 2d CFT to trace over left- or right-moving modes.

There are a few questions that our work suggests. One of them is the extension of our calculations to more general boundary states in arbitrary CFT's.  What we envision is a calculation in an arbitrary BCFT using the more abstract description presented, for example, in \cite{Cardy:2004hm}. Many of the key ingredients in our computation are available in the general case. We hope to return to this question in the future.  It is plausible that a universal formula, such as the one for the entanglement entropy of a segment ($S \sim c/3$) could emerge, although conceivably
there should be some dependence or more than just the central charge, as we have seen explicitly in the case of the free boson CFT.

It would also be interesting, and we hope to return to this question soon, to discuss the boundary states that appear in string theory with all its decorations. Extending our calculations to string theory D-branes should help us make a bridge to p-branes in supergravity and potentially gain some insight into the Bekenstein-Hawking entropy of such configurations.

\section*{Acknowledgments}
We would like to thank  P. Amore, A. Bhattacharyya, C. Terrero-Escalante and G. Wong. We are particularly grateful to T. Takayanagi for useful discussions and insightful comments. L.A.P.Z. thanks the Abdus Salam ICTP, Italy for hospitality at various stages of this project. This work is partially
supported by Department of Energy under grant DE-
FG02-95ER40899. N. Quiroz thanks to  Facultad de Ciencias, Universidad de Colima for support.

\begin{appendix}{Conventions and technical details}\label{Sec:app}
We work in units of $\alpha'=2$. The $d=1$ closed string field is
\be
X(\tau,\sigma)=x + 2 p \tau +i
\sum_{n \neq 0}\frac{1}{n}\left( \alpha_n e^{-in(\tau+\sigma)} + \tilde \alpha_n e^{-in(\tau-\sigma)} \right).
\ee
The closed string hamiltonian is $H=H_L+H_R$
\bea
\label{eq:ch}
H_L&=& \pi \alpha_0^2 + 2\pi\sum_{n=1}^\infty \alpha_{-n}\alpha_n -  2\pi\frac{c}{24}, \nonumber \\
H_R&=& \pi \tilde{\alpha}_0^2 +2\pi  \sum_{n=1}^\infty \tilde{\alpha}_{-n}\tilde{\alpha}_n -2\pi \frac{\tilde c}{24}.
\eea
where in our specific case $c=\tilde c =1$.

In the case of compactification $X(\tau,\sigma+2\pi)\sim X(\tau,\sigma)+2\pi R$ and the left and right momenta are not in general the same. They are now

\be
P_R \equiv \tilde\alpha_0=\frac{m}{R}-\frac{kR}{2}, \qquad P_L \equiv   \alpha_0=\frac{m}{R}+\frac{kR}{2}\,,
\ee
where $m$ is the momentum and $k$ is the winding quantum numbers, respectively. The ground state is characterized by these two integers: $m$ and $k$. Hereafter $\ket{\frac{m}{R}, \frac{kR}{2}}$ denotes the  oscillator vacuum with zero-mode parameters $m$ and $k$.

 The closed string Hamiltonian is
\be
H_c= 2\pi\left(\frac{m^2}{R^2} + \frac{R^2k^2}{4}\right) +2\pi\sum_n(\alpha_{-n}\alpha_n + \tilde \alpha_{-n} \tilde \alpha_n ) -\frac{2\pi}{12}.
\ee

The tree-level amplitude describing the interaction of two boundary states is
\be
\label{eq:treeamp}
Z_c(l)=\bra{B}e^{-2 \pi l H}\ket{B}\,,
\ee
where $l$ is the length of the cylinder bounded by the boundary states.

For Neumann boundary state
\be
Z_c(l)= {\cal N}^2\, \frac{1}{\eta(2il)}  \;\;\;\;\mbox{with}\;\;\; {\cal N}= \frac{1}{\sqrt{2}}\,,
\ee
while for Dirichlet boundary state the tree amplitude receives a contribution of the zero modes
\be
Z_c(l)= {\cal N}^2\,e^{-\frac{l}{8\pi}}\frac{1}{\sqrt{2l}} \ \frac{1}{\eta(2il)}  \;\;\;\; \mbox{with}\;\;\; {\cal N}= 1\,.
\ee
The reduced density matrix $\rho_A$  for the Neumann boundary state is given as

\be
\begin{split}
\label{eq:rA}
\rho_{A}& = \frac{1}{{\cal Z}}\displaystyle\sum_{\vec{s}} \bra{\vec{\tilde{s}}}e^{-\epsilon H}\ket{B}\bra{B}e^{-\epsilon H}
\ket{\vec{\tilde{s}}}\\
&=\frac{{\cal N}^2}{\cal Z} e^{\pi \epsilon/3}\displaystyle\sum_{\vec{\tilde s}}  \prod_{t=1}^{\infty}e^{-8\pi \epsilon t s_t}\frac{1}{s_t}\left(\frac{\alpha_{-t}}{\sqrt{t}}\right)^{s_t}\ket{0}\bra{0} \left(\frac{\alpha_{t}}{\sqrt{t}}\right)^{s_t}.
\end{split}
\ee
The above expression for the density matrix lead us, naturally, to the concept of boundary entropy which is obtained in the large $\epsilon$ limit \cite{Affleck:1991tk}.  Recall that the boundary entropy  of the boundary state $\ket{B}$ is defined as
\be
s=\mbox{ln}g\,,
\ee
with $g=\braket{0}{B}={\cal N}$. The reason we do not see explicitly the boundary entropy in our calculations is because we work in the opposite limit, namely, $\epsilon\to 0$, where the contribution of the ground state need not dominate the density matrix.
This matrix is diagonal and
\be
\label{eq:Ntrace}
\mbox{Tr}\rho_A = \frac{{\cal N}^2}{{\cal A}}e^{\frac{\pi \epsilon}{3}}\prod_{l=1}^{\infty}\frac{1}{1-e^{-8 \pi l \epsilon}}=  \frac{{\cal N}^2}{{\cal A}}\frac{1}{\eta(4i\epsilon)}\,,
\ee
 and  since $Tr \rho= Tr \rho_A =1$, then ${\cal A}= \frac{{\cal N}^2}{\eta(4i\epsilon)}$.
For the R\'enyi entropy we will need
\be
\label{eq:Ntracen}
\mbox{Tr}\rho_A^n = \frac{\eta^n(4i\epsilon)}{\eta(4i\epsilon n)}\,.
\ee
We know write $Tr \rho_A^n$ in a different way. For that recall Eq.(\ref{eq:treeamp}), then
\be\label{eq:Gtracen}
\mbox{Tr}\rho_A^n \propto \frac{Z_c(2\epsilon n)}{Z_c(2\epsilon)^n}  = \frac{\bra{B}e^{-2 \epsilon n H}\ket{B}}{\left({\bra{B}e^{-2\epsilon H}\ket{B}}\right)^n}\,.
\ee
For Dirichlet brane $\mbox{Tr}\rho_A^n$ is defined as above but now there is a contribution of the zero-modes to the density matrix.
Its trace is
\be
\label{eq:Dtrace}
\mbox{Tr}\rho_{A}
= e^{-\frac{a^2}{4\pi \epsilon}}  \frac{{\cal N}^2}{\cal A}
\frac{1}{2\epsilon }\frac{1}{\eta(4i\epsilon)}\,,
\ee
from where  ${\cal A}$ is fixed. The trace of $\rho_A^n$ has the same form as  equation (\ref{eq:tracen}) for the oscillators. Together with the zero mode contribution one has
\be
\label{eq:Dtracen}
\mbox{Tr}\rho_A^n=\frac{\sqrt{4\epsilon}^n}{\sqrt{4\epsilon n}}
\frac{\eta^n(4i\epsilon)}{\eta(4i\epsilon n)}.
\ee

The prescription given for the non-compact case works also for the compact case. For the Neumann state

$$
{\cal A} = {\cal N}^2\sum_k e^{-(2\epsilon) k^2 R^2}\frac{1}{\eta(4i\epsilon)}\,.
$$
Again by  Eq.(\ref{eq:Gtracen}) one can obtain

$$
\mbox{Tr}\rho_A^n = \frac{\vartheta_3\left( 0| i\epsilon  n R^2  \right)}{\eta(4i\epsilon n)}\frac{\eta^n(4i\epsilon)}{\vartheta_3^n\left(0|i \epsilon  R^2  \right)}\,.
$$
For the compact Dirichlet state
$$
\mbox{Tr}\rho_A^n =\frac{\vartheta_3(0|\frac{4i\epsilon n}{R^2})}{\eta(4i\epsilon n)}\frac{\eta^n(4i\epsilon)}{\vartheta_3^n(0,\frac{4i\epsilon)}{R^2}}\,.
$$

\end{appendix}


\bibliographystyle{JHEP}
\bibliography{entanglement}

\end{document}